\documentclass[12pt]{iopart}

\usepackage{iopams}
\usepackage{amssymb}
\eqnobysec
\newtheorem{theorem}{Theorem}[section]
     \newtheorem{lemma}[theorem]{Lemma}
     
     \newtheorem{corollary}[theorem]{Corollary}
     \newtheorem{definition}[theorem]{Definition}
     \newenvironment{proof}[1][Proof.]{\begin{trivlist}
     \item[\hskip \labelsep {\bfseries #1}]}{\end{trivlist}}

\begin{document}

\title[]{On the bound states of the Dirac equation in the extreme Kerr metric}

\author{D. Batic}
\address{Department of Mathematics, University of Los Andes, AA 4976 Bogota, Colombia}
\ead{dbatic@uniandes.edu.co,batic@itp.phys.ethz.ch}
\author{M. Nowakowski}
\address{Department of Physics, University of Los Andes, AA 4976 Bogota, Colombia}
\ead{mnowakos@uniandes.edu.co}
\begin{abstract}
We study the eigenvalues of the angular equation arising after the
separation of the Dirac equation in the extreme Kerr metric. To this
purpose a self-adjoint holomorphic operator family associated to
this eigenvalue problem is considered. We show that the eigenvalues
satisfy a first order nonlinear differential equation with respect
to the black hole mass and we solve it. Finally, we prove that there
exist no bound states for the Dirac equation in the aforementioned
metric.
\end{abstract}

\maketitle

\section{Introduction}\label{sec:1}
The last ten years have been characterized by an increasing interest
in studying the behavior of Dirac particles in the geometry of an
extreme black hole \cite{BEL,SC,MO,BA1}. The main picture arising
from the aforementioned studies is that under certain conditions on
the physical parameters the Dirac equation in the
Reissner-Nordstr\"{o}m, extreme Kerr and extreme Kerr-Newman metrics
might admit the existence of bound state solutions. In turn, this
leads to the tempting interpretation of such systems made of an
extreme black hole plus a fermion as a new kind of atomic system
with an extreme charged or uncharged black hole as its nucleus and
around an
electronic cloud.\\
A still open problem is to understand if such results could be
relevant for astrophysics. The present work represents a
contribution in this direction. Despite the common belief that the
formation of an extreme Kerr black hole (EKBH) is of only academic
interest it is our opinion that we cannot a priori exclude that
EKBH's play no role in astrophysics. For instance, relativistic
Dyson rings admit a continuous transition to an EKBH \cite{ANS}.
Studies about the existence of such rings based on numerical
computations can be found in \cite{WON,ERI}. Moreover, relativistic
Dyson rings could emerge from astrophysical scenarios like stellar
core-collapses with high angular momentum \cite{TH} or they might
simply be present in central regions of galaxies. Finally, it has
been recently proved that the only possible candidate for a black
hole limit for stationary and axisymmetric, uniformly rotating
perfect fluid bodies with a cold equation of state as well as for
isentropic stellar models with a non-zero temperature is the EKBH
\cite{MEI}. Hence, it appears reasonable to study the Dirac equation
in
the geometry of an extreme black hole.\\
In what follows we restrict our attention to extreme Kerr black
holes. Although it is not the most general model of the exterior
region of a black hole we can analyze theoretically, it represents
indeed the most realistic model in astrophysics since in general
black holes are embedded in environments that are rich in gas and
plasma and, consequently any net charge is neutralized by the
ambient plasma. The main problem connected with the bound states for
the Dirac equation in the extreme Kerr and Kerr-Newman metrics
\cite{SC,MO} is that an energy eigenvalue $\omega$ has to satisfy a
complicated set of conditions. Let us now consider an EKBH with mass
$M$, angular momentum per unit mass $a=M$. Let $m_e$ and
$k=\pm1/2,\pm3/2,\cdots$ be, the mass and the azimuthal quantum
number of a spin $1/2$ particle, respectively. Then, according to
\cite{SC} the following conditions are necessary
\begin{equation}\label{cond1}
\fl 2M\omega+k=0,\qquad  m_e^2-\omega^2>0,\qquad
\lambda_{nk}^2+M^2\left(m^2_e-4\omega^2\right)>\frac{1}{4}
\end{equation}
for $\omega$ to be an eigenvalue. Here, $\lambda_{nk}\in\mathbb{R}$
with $n\in\mathbb{Z}\backslash\{0\}$ denotes an eigenvalue of the
angular problem arising from the Dirac equation after separation of
variable by means of the Chandrasekhar ansatz \cite{CH}. However,
the solvability of the above system is not yet sufficient for the
existence of an energy eigenvalue. \cite{SC} showed that if in
addition either
\begin{equation}\label{cond2}
\fl \frac{Mm_e\omega}{\sqrt{m_e^2-\omega^2}}+\lambda_{nk}=0,\qquad
\frac{M(m_e^2-2\omega^2)}{\sqrt{m_e^2-\omega^2}}+\sqrt{\lambda_{nk}^2+M^2\left(m^2_e-4\omega^2\right)}=0
\end{equation}
or
\begin{equation}\label{cond3}
\fl
N+\frac{M(m_e^2-2\omega^2)}{\sqrt{m_e^2-\omega^2}}+\sqrt{\lambda_{nk}^2+M^2\left(m^2_e-4\omega^2\right)}=0,\quad
\mbox{for some positive integer $N$}
\end{equation}
holds, then the solvability of the system (\ref{cond1}) with
(\ref{cond2}) or (\ref{cond3}) is sufficient for the existence of an
eigenvalue $\omega$. It is not clear that for given data of the
black hole and the particle the system (\ref{cond1}) with
(\ref{cond2}) or (\ref{cond3}) is satisfied. Our present work is
aimed to answer this question. There is one disturbing point
concerning the first equation entering in (\ref{cond1}), namely the
supposed energy eigenvalues do not depend on both quantum numbers
$n$ and $k$ but only on $k$. From a physical point of view this is
strange since a simple analogy with the hydrogen atom would suggest
that the energy eigenvalues should indeed depend on both quantum
numbers $n$ and $k$.\\ The rest of the paper is organized as
follows. In section~\ref{sec:2} we shortly derive the Dirac Equation
in the EKBH. Section~\ref{sec:3} is devoted to derive and to solve a
nonlinear ODE for the eigenvalues $\lambda$ with respect to the
black hole mass parameter. In section~\ref{sec:4} we construct a
quasi-linear PDE for the eigenvalues with respect to the energy of
the particle and the mass of the black hole. Moreover, we derive a
formal power solution for $\lambda$. Finally, in section~\ref{sec:5}
we show that the solution set of the system (\ref{cond1}) with
(\ref{cond2}) or (\ref{cond3}) is empty. This result implies that
there are no bound state solutions for the Dirac equation in the
extreme Kerr metric.

\section{The Dirac equation in the extreme Kerr metric}\label{sec:2}
In Boyer-Lindquist coordinates $(t,r,\vartheta,\varphi)$ with $r>0$,
$0\leq\vartheta\leq\pi$, $0\leq\varphi<2\pi$ the extreme Kerr metric
 \cite{BA1} is given by
\begin{equation}\label{KN}
\fl \rmd s^{2}= \left(1-\frac{2Mr}{\Sigma}\right)\rmd
t^{2}+\frac{4M^{2}r\sin^{2}\vartheta}{\Sigma}\rmd t
\rmd\varphi-\frac{\Sigma}{\Delta}\rmd r^2-\Sigma ~\rmd\vartheta^{2}
-(r^2+M^2)^{2}\sin^{2}{\vartheta}\frac{\widetilde{\Sigma}}{\Sigma}\rmd\varphi^{2}
\end{equation}
with
\[
\fl \Sigma:=\Sigma(r,\vartheta)=r^2+M^2\cos^{2}\vartheta, \qquad
\Delta:=\Delta(r)=(r-M)^2
\]
and
\[
\fl
\widetilde{\Sigma}:=\widetilde{\Sigma}(r,\vartheta)=1-M^{2}\gamma^{2}(r)\sin^{2}{\vartheta},\quad
\gamma(r):=\frac{r-M}{r^2+M^2}
\]
where $M$ is the mass of a spinning black hole with angular momentum
$J=M^2$. Since the equation $\Delta=0$ has a double root at
$r_{0}:=M$ the Cauchy horizon and the event horizon coincide.\\
In the following we consider a spin-$\frac{1}{2}$ particle with mass
$m_e$ and charge $e$ in the extreme Kerr background. The behavior of
such a particle is governed by the Dirac equation, a linear system
of four coupled partial differential equations. In the extreme Kerr
metric the Dirac equation can be easily obtained from the results in
\cite{FEL} by setting the Kerr parameter $a$ equal to the mass $M$
of the black hole and it has the form
\begin{equation} \label{Dirac}
\fl \left(\mathcal{R}+\mathcal{A}\right)\Psi=0
\end{equation}
where
\begin{equation*} \label{2}
\fl \mathcal{R}=\left( \begin{array}{cccc}
                            \rmi m_{e}r&0&\sqrt{\Delta}\mathcal{D}_{+}&0\\
                            0&-\rmi m_{e}r&0&\sqrt{\Delta}\mathcal{D}_{-}\\
                            \sqrt{\Delta}\mathcal{D}_{-}&0&-\rmi m_{e}r&0\\
                             0&\sqrt{\Delta}\mathcal{D}_{+}&0&\rmi m_{e}r
                            \end{array} \right),
\end{equation*}
\begin{equation*} \label{3}
\fl \mathcal{A}=\left( \begin{array}{cccc}
                            -Mm_{e}\cos{\vartheta}&0&0&\mathcal{L}_{+}\\
                            0&Mm_{e}\cos{\vartheta}&-\mathcal{L}_{-}&0\\
                            0&\mathcal{L}_{+}&-Mm_{e}\cos{\vartheta}&0\\
                            -\mathcal{L}_{-}&0&0&Mm_{e}\cos{\vartheta}
                            \end{array} \right)
\end{equation*}
with $\mathcal{D}_{\pm}$ and $\mathcal{L}_{\pm}$ defined by
\begin{eqnarray*}
&&\fl\mathcal{D}_{\pm}=\frac{\partial}{\partial r}\mp\frac{1}{\Delta}\left[(r^2+M^2)\frac{\partial}{\partial t}+M\frac{\partial}{\partial\varphi}\right],\label{DPM}\\
&&\fl\mathcal{L}_{\pm}=\frac{\partial}{\partial
\vartheta}+\frac{1}{2}\cot{\vartheta}\mp
\rmi\left(M\sin{\vartheta}\frac{\partial}{\partial
t}+\csc{\vartheta}\frac{\partial}{\partial\varphi}\right)\label{LPM}.
\end{eqnarray*}
By rearranging (\ref{Dirac}) we can write the Dirac equation in
Hamiltonian form
\begin{equation}\label{Dirac_Ham}
\fl \rmi\partial_t\Psi=H\Psi
\end{equation}
where $H$ is a first order $4\times 4$ matrix differential operator
acting on spinors $\Psi$ on hypersurfaces $t=\mbox{const}.$
Similarly as in \cite{FEL} we can construct a positive scalar
product
\begin{equation}\label{scalar_product}
\fl \langle\Psi|\Phi\rangle=\int_{M}^{+\infty}\,\rmd
r\int_{0}^{\pi}\,\rmd\vartheta\int_{0}^{2\pi}\,\rmd\varphi~\overline{\Psi}(t,r,\vartheta,\varphi)
\Phi(t,r,\vartheta,\varphi)\frac{r^2+M^2}{\Delta}
\end{equation}
where $\overline{\Psi}$ denotes the complex conjugated transposed
spinor. In the present work we are interested in time periodic
solutions
\[
\fl \Psi(t,r,\vartheta,\varphi)=\rme^{-\rmi\omega
t}\Psi_{0}(r,\vartheta,\varphi)
\]
of the Dirac equation (\ref{Dirac}) where $\omega\in\mathbb{R}$ and
$\Psi_0$ is normalizable, that is
$\langle\Psi|\Phi\rangle=\langle\Psi_0|\Phi_0\rangle=1$. Notice that
 if such a solution exists, then $\omega$ is an eigenvalue of $H$ for
the eigenspinor $\Psi_0$ and $\omega$
represents the particle energy of the bound state $\Psi$.\\
By means of the Chandrasekhar ansatz \cite{CH1}
\[
\fl \Psi_0(r,\vartheta,\varphi)=\rme^{-\rmi k\varphi}\left(
\begin{array}{c}
               f_{1}(r)g_{1}(\vartheta)\\
               f_{2}(r)g_{2}(\vartheta)\\
               f_{2}(r)g_{1}(\vartheta)\\
               f_{1}(r)g_{2}(\vartheta)
\end{array} \right),\quad k\in\{\pm 1/2,\pm 3/2,\cdots\}
\]
the Dirac equation decouples into the equations
\[
\fl \mathcal{R}\Psi=\lambda\Psi,\qquad \mathcal{A}\Psi=-\lambda\Psi
\]
with separation parameter $\lambda\in\mathbb{R}$. Finally, defining
\[
\fl f(r):=\left(
\begin{array}{c}
               f_{1}(r)\\
               f_{2}(r)
\end{array} \right),\qquad g(\vartheta):=\left(
\begin{array}{c}
               g_{1}(\vartheta)\\
               g_{2}(\vartheta)
\end{array} \right)
\]
the Dirac equation can be separated into a radial part
\begin{eqnarray}
&&\fl\left( \begin{array}{cc}
     (r-M)\frac{d}{dr}+\rmi\frac{V(r)}{r-M}&\rmi m_{e}r-\lambda\\
     -\rmi m_{e}r-\lambda&(r-M)\frac{d}{dr}-\rmi\frac{V(r)}{r-M}
           \end{array} \right)f(r)=0, \label{radial}\\
&&\fl\left( \begin{array}{cc}
     \frac{d}{d\vartheta}+\frac{1}{2}\cot{\vartheta}-Q(\vartheta) & \lambda-Mm_{e}\cos{\vartheta}\\
                \lambda+Mm_{e}\cos{\vartheta} & -\frac{d}{d\vartheta}-\frac{1}{2}\cot{\vartheta}-Q(\vartheta)
           \end{array} \right)g(\vartheta)=0 \label{angular}
\end{eqnarray}
where
\[
\fl V(r)=\omega(r^2+M^2)+\kappa M, \qquad
Q(\vartheta)=M\omega\sin{\vartheta}+k\csc{\vartheta}.
\]
As in \cite{SC} we introduce the following definition
\begin{definition}
We say that $\omega\in\mathbb{R}$ is an energy eigenvalue of
(\ref{Dirac}) if there exists a $\lambda\in\mathbb{R}$ and
nontrivial solutions $f$ of (\ref{radial}) and $g$ of
(\ref{angular}) satisfying the normalization conditions
\begin{equation}\label{integrability_cond}
\fl \int_{M}^{+\infty}\,\rmd r
\frac{r^2+M^2}{\Delta}~|f(r)|^2=1,\qquad
\int_{0}^{\pi}\,\rmd\vartheta \sin{\vartheta}~|g(\vartheta)|^2=1.
\end{equation}
\end{definition}

\section{An ordinary differential equation for the eigenvalues\label{sec:3}
$\lambda$}\label{sec:3} By means of the transformation
\[
\fl \widetilde{g}(\vartheta):=\left(
\begin{array}{c}
               \widetilde{g}_{1}(\vartheta)\\
               \widetilde{g}_{2}(\vartheta)
\end{array} \right)=\sqrt{\sin{\vartheta}}~g(\vartheta)
\]
the angular equation takes the form
\begin{equation}\label{system_ODE}
\fl (\mathfrak{U}~\widetilde{g})(\vartheta):=\left(
\begin{array}{cc}
       0&\frac{\rmd}{\rmd\vartheta}+\frac{k}{\sin{\vartheta}}\\
      -\frac{\rmd}{\rmd\vartheta}+\frac{k}{\sin{\vartheta}}&0
                     \end{array} \right)\widetilde{g}
+M\left( \begin{array}{cc}
       -m_{e}\cos{\vartheta}&\omega\sin{\vartheta}\\
       \omega\sin{\vartheta}&m_{e}\cos{\vartheta}
                    \end{array}
                     \right)\widetilde{g}=\lambda\widetilde{g}
\end{equation}
with $\vartheta\in(0,\pi)$. It is straightforward to check that the
solutions $\widetilde{g}_1$ and $\widetilde{g_2}$ of
(\ref{system_ODE}) have the following useful property
\begin{equation}\label{proprieta}
\fl \widetilde{g}_1(\pi-\vartheta)=\widetilde{g}_2(\vartheta),\qquad
\widetilde{g}_2(\pi-\vartheta)=\widetilde{g}_1(\vartheta).
\end{equation}
We can associate the minimal operator $\mathcal{A}_0$ to the formal
differential expression $\mathfrak{U}$ acting in the Hilbert space
$\mathcal{H}:=L_{2}((0,\pi)^{2},\mathbb{C}^2)$ of square integrable
vector functions with respect to the scalar product
\[
\fl
\left(\widetilde{g}_1,\widetilde{g}_2\right)=\int_{0}^{\pi}\,\rmd\vartheta
~{\widetilde{g}_{2}}^{*}~\widetilde{g}_1,\qquad
\widetilde{g}_1,\widetilde{g}_2\in\mathcal{H}.
\]
The operator $\mathcal{A}_0$ given by
$D(\mathcal{A}_0)=\mathcal{C}_{0}^{\infty}((0,\pi),\mathbb{C}^2)^{2}$
and $\mathcal{A}_0\widetilde{g}:=\mathfrak{U}\widetilde{g}$ for
$\widetilde{g}\in D(\mathcal{A}_0)$ is densely defined and closable.
Moreover, since the formal differential operator $\mathfrak{U}$ is
in the limit point case at $0$ and $\pi$ it follows that
$\mathcal{A}_0$ is essentially self-adjoint. In the following we
denote the closure of $\mathcal{A}_0$ by $\mathcal{A}$. To indicate
the dependence of the angular operator $\mathcal{A}$ and its
eigenvalues $\lambda$ on the parameter $M$ we use the notation
$\mathcal{A}(M)$ and $\lambda(M)$.\\
According to \cite{WE1} (Thm.$5.8$) the domain of $\mathcal{A}(0)$
is given by
\[
\fl
D(\mathcal{A})=\{\widetilde{g}\in\mathcal{H}~:~\mbox{$\widetilde{g}$
is absolutely continuous and
$\mathcal{A}(0)\widetilde{g}\in\mathcal{H}$}\}.
\]
Since $\mathcal{A}(M)=\mathcal{A}(0)+T(M)$ with the bounded
multiplication operator
\[
\fl T(M)=M\left( \begin{array}{cc}
       -m_{e}\cos{\vartheta}&\omega\sin{\vartheta}\\
       \omega\sin{\vartheta}&m_{e}\cos{\vartheta}
                    \end{array}
                     \right)
\]
its domain of definition $D(\mathcal{A})$ is independent of
$M\in\mathbb{C}$ (see \cite{KATO} Chap. IV, $\S$1, Thm.1.1).
Moreover, for $M\in\mathbb{R}$ it results that $T(M)$ is a symmetric
perturbation of $\mathcal{A}(0)$ and Thm.4.10, Chap. V, $\S$4 in
\cite{KATO} implies that $\mathcal{A}(M)$ is self-adjoint. According
to the classification in \cite{KATO} (Chap. VII, $\S$3)
$\mathcal{A}(M)$ forms a self-adjoint holomorphic operator family of
type (A) in the variable $M\in\mathbb{C}$. Further, the spectrum of
$\mathcal{A}(0)$ is discrete and consists of simple eigenvalues
given by Lemma~3.3, Chap.3, $\S$1.2 in \cite{MON}
\begin{equation}\label{initial_condition}
\fl \lambda_{n,k}
(0)=\mbox{sign(n)}\left(|k|-\frac{1}{2}+|n|\right),\qquad
n\in\mathbb{Z}\backslash\{0\}.
\end{equation}
This means that $\mathcal{A}(0)$ has compact resolvent and Thm.2.4,
Chap. V, $\S$2 in \cite{KATO} yields that $\mathcal{A}(M)$ has
compact resolvent for all $M\in\mathbb{C}$. This implies that the
eigenvalues $\lambda_{n,k}=\lambda_{n,k}(M)$,
$n\in\mathbb{Z}\backslash\{0\}$ of $\mathcal{A}(M)$ are simple and
depend holomorphically on $M$. Moreover, the first derivative of
$\mathcal{A}$ with respect to $M$ is given by
\[
\fl \frac{d\mathcal{A}}{dM}=\left( \begin{array}{cc}
       -m_{e}\cos{\vartheta}&\omega\sin{\vartheta}\\
       \omega\sin{\vartheta}&m_{e}\cos{\vartheta}
                    \end{array}
                     \right)
\]
which yields the following estimates for the growth rate of the
eigenvalues (see \cite{KATO} Chap. VII, $\S$3.4)
\[
\fl
\left|\frac{d\lambda_{n,k}}{dM}\right|=\left|\left|\frac{d\mathcal{A}}{dM}\right|\right|\leq\max{\{|m_e|,|\omega|\}}.
\]
Here, $\|\cdot\|$ denotes the operator norm of a $2\times 2$ matrix.
In addition, Thm.4.10, Chap. V, $\S$3 in \cite{KATO} implies that
\[
\fl
\min_{n\in\mathbb{Z}\backslash\{0\}}\left|\lambda_{n,k}-\lambda_{n,k}(0)\right|\leq\|T(M)\|\leq\max{\{|m_e|,|\omega|\}}
\]
for each eigenvalue $\lambda_{n,k}$ of $\mathcal{A}(M)$. Finally, by
interchanging the components of $\widetilde{g}(\vartheta)$ it is
easy to check that $\lambda_{n,k}$ is an eigenvalue of $\mathcal{A}$
if and only if $-\lambda_{-n,-k}$ is an eigenvalue of $\mathcal{A}$
with $k$, $M$, and $m_e$ replaced by $-k$, $-M$ and $-m_e$,
respectively. Since the eigenvalues depend holomorphically on $M$
the following identity holds
\[
\fl \lambda_{n,k}(\omega,m_e;M)=-\lambda_{-n,-k}(\omega,-m_e;-M)
\]
for all $M\in\mathbb{R}$. Therefore, without loss of generality we
can always restrict our attention to the case $k\geq 1/2$.
\begin{theorem}\label{Equation_for_lambda}
For fixed $k$, $\omega$ and $m_e$ the eigenvalue $\lambda_{n,k}$ of
$\mathcal{A}$ satisfies the first order nonlinear separable
differential equation
\begin{equation}\label{ODE}
\fl \frac{d\lambda_{n,k}}{dM}=2(M\omega+k)\frac{2\omega\lambda_{n,k}
-m_e}{4\lambda_{n,k}^2-1}
\end{equation}
where $\lambda_{n,k}(0)$ is given by (\ref{initial_condition}).
\end{theorem}
\begin{proof}
For simplicity in notation we omit in the following the indices $n$
and $k$ of $\lambda$. Let $\widetilde{g}$ be that eigenfunction of
$\mathcal{A}$ for the eigenvalue $\lambda$ which is normalized by
the condition $(\widetilde{g},\widetilde{g})=1$. Introducing the
functions
\[
\fl
U(\vartheta):=\widetilde{g}_1^2(\vartheta)+\widetilde{g}_2^2(\vartheta),\qquad
V(\vartheta):=\widetilde{g}_2^2(\vartheta)-\widetilde{g}_1^2(\vartheta),\qquad
W(\vartheta):=2\widetilde{g}_1(\vartheta)\widetilde{g}_2(\vartheta)
\]
and employing (\ref{system_ODE}) it can be easily checked that $U$,
$V$, and $W$ satisfy the following system of ODEs
\begin{eqnarray}
\fl U^{'}(\vartheta)&=&-2f(\vartheta)V(\vartheta)+Mm_{e}\cos{\vartheta}~W(\vartheta),\qquad f(\vartheta)=M\omega\sin{\vartheta}+\frac{k}{\sin{\vartheta}},\label{uno}\\
\fl V^{'}(\vartheta)&=&-2f(\vartheta)U(\vartheta)+2\lambda W(\vartheta),\label{due}\\
\fl W^{'}(\vartheta)&=&2Mm_{e}\cos{\vartheta}~U(\vartheta)-2\lambda
V(\vartheta).\label{tre}
\end{eqnarray}
From analytic perturbation theory (see \cite{KATO}, Chap. VII,
$\S$3.4) we have
\begin{equation}\label{primaequa}
\fl
\frac{d\lambda}{dM}=\left(\frac{d\mathcal{A}}{dM}~\widetilde{g},\widetilde{g}\right)=\int_{0}^{\pi}\,\rmd\vartheta~
\widetilde{g}^{*}(\vartheta)\left( \begin{array}{cc}
       -m_{e}\cos{\vartheta}&\omega\sin{\vartheta}\\
       \omega\sin{\vartheta}&m_{e}\cos{\vartheta}
                    \end{array}
                     \right)\widetilde{g}(\vartheta)=m_{e}I_1+\omega I_2
\end{equation}
with
\[
\fl
I_1=\int_{0}^{\pi}\,\rmd\vartheta\cos{\vartheta}~V(\vartheta),\qquad
I_2=\int_{0}^{\pi}\,\rmd\vartheta\sin{\vartheta}~W(\vartheta).
\]
In addition, from Lemma 1 in \cite{BA} the following estimates hold
\begin{equation}\label{estimate}
\fl |U(\vartheta)|,|V(\vartheta)|,|W(\vartheta)|\leq
C\sin^{2k}{\vartheta}
\end{equation}
with some constant $C>0$. Since without loss of generality $k$ can
be assumed positive, it results that $U$, $V$ and $W$ vanish at
$\vartheta=0$ and $\vartheta=\pi$. If we multiply (\ref{due}) by
$\sin{\vartheta}$, integrate by parts and take into account that
$\int_{0}^{\pi}\,\rmd\vartheta~U(\vartheta)=1$ we obtain
\begin{equation}\label{primero}
 \fl 2(M\omega+k)=I_1 +2\lambda
I_2+2M\omega I_3,\qquad
I_3:=\int_{0}^{\pi}\,\rmd\vartheta~\cos^{2}{\vartheta}~U(\vartheta).
\end{equation}
If we multiply (\ref{tre}) by $\cos{\vartheta}$ and integrate by
parts we get
\begin{equation}\label{secundo}
 \fl 2\lambda I_1 +I_2=2Mm_e I_3.
\end{equation}
The next step consists in computing the integral entering on the
l.h.s. in (\ref{primero}) and (\ref{secundo}). By means of
(\ref{due}) we can rewrite (\ref{uno}) and (\ref{tre}) in terms of
the functions $U$ and $V$ and their first derivatives as follows
\begin{equation}\label{uno1}
\fl
U^{'}(\vartheta)+2f(\vartheta)V(\vartheta)=\frac{Mm_{e}}{2\lambda}\cos{\vartheta}\left(V^{'}(\vartheta)+2f(\vartheta)U(\vartheta)\right),
\end{equation}
\begin{equation}\label{due1}
\fl
\left(V^{'}(\vartheta)+2f(\vartheta)U(\vartheta)\right)^{'}=4Mm_{e}\lambda\cos{\vartheta}U(\vartheta)-4\lambda^2
V(\vartheta).
\end{equation}
If we derivate (\ref{uno1}) once with respect to $\vartheta$ and
make use of (\ref{due1}) we obtain
\begin{equation*}
\fl
\left(U^{'}(\vartheta)+2f(\vartheta)V(\vartheta)\right)^{'}=-\frac{Mm_e}{2\lambda}\sin{\vartheta}\left(V^{'}(\vartheta)+2f(\vartheta)U(\vartheta)\right)+
\end{equation*}
\begin{equation}\label{equa}
+2Mm_{e}\cos{\vartheta}\left(Mm_{e}\cos{\vartheta}U(\vartheta)-\lambda
V(\vartheta)\right).
\end{equation}
Notice that
\[
\fl
\int_{0}^{\pi}\,\rmd\vartheta~\left(U^{'}(\vartheta)+2f(\vartheta)V(\vartheta)\right)^{'}=\left.\left(U^{'}(\vartheta)+2f(\vartheta)V(\vartheta)\right)\right|_0^\pi=\left.Mm_e\cos{\vartheta}~W(\vartheta)\right|_0^\pi=0
\]
because of (\ref{estimate}). Hence, if we integrate (\ref{equa}) on
the interval $(0,\vartheta)$ we get
\begin{equation}\label{equazia}
\fl 2M(2m_e\lambda-\omega)I_3+(1-4\lambda^2)I_1=2(M\omega+k).
\end{equation}
Equations (\ref{primero}), (\ref{secundo}) and (\ref{equazia}) form
a system for the unknowns $I_1$, $I_2$ and $I_3$. We solve it and we
find
\begin{equation}\label{integrals}
\fl I_1=2\frac{M\omega+k}{1-4\lambda^2},\qquad I_2=-2\lambda
I_1,\qquad I_3=0.
\end{equation}
Insertion of (\ref{integrals}) into (\ref{primaequa}) gives
(\ref{ODE}). This completes the proof.\hspace{0.5cm}$\square$
\end{proof}

\begin{theorem}\label{solution}
For fixed $k$, $\omega\neq 0$ and $m_e$ there exists a unique
solution of (\ref{ODE}) subjected to the initial condition
(\ref{initial_condition}). The solution in implicit form is given by
\begin{equation*}
\fl
\omega\left(\lambda^2_{n,k}(M)-\lambda^2_{n,k}(0)\right)+m_e\left(\lambda_{n,k}(M)-\lambda_{n,k}(0)\right)-\frac{\omega^2-m_e^2}{2\omega}
\ln{\left(\frac{2\omega\lambda_{n,k}(M)-m_e}{2\omega\lambda_{n,k}(0)-m_e}\right)}=
\end{equation*}
\begin{equation}\label{general_solution}
=M\omega^2 (M\omega+2k).
\end{equation}
\end{theorem}
\begin{proof}
For simplicity in notation we omit in the following the indices $n$
and $k$ of $\lambda$. Since (\ref{ODE}) is separable the solution
(\ref{general_solution}) satisfying the initial condition
(\ref{initial_condition}) can be obtained by computing the integrals
entering in the following expression
\[
\fl
\int_{\lambda(0)}^{\lambda(M)}\,\rmd\lambda~\frac{4\lambda^2-1}{2\omega\lambda-m_e}=2\int_{0}^{M}\,\rmd
M(M\omega+k).\hspace{0.5cm}\square
\]
\end{proof}
\begin{corollary}\label{special_solution}
For fixed $k$, $m_e$ and $\omega=0$ the solution of (\ref{ODE})
subjected to the initial condition (\ref{initial_condition}) is
given by
\begin{equation*}
\fl
\lambda_{n,k}(M)=\frac{1+\left(-3c_{n,k}(M)+\sqrt{9c^2_{n,k}(M)-1}\right)^{2/3}}{2\sqrt[3]{-3c_{n,k}(M)+
\sqrt{9c^2_{n,k}(M)-1}}},
\end{equation*}
\begin{equation*}
\fl c_{n,k}(M)=2m_e
kM+\lambda_{n,k}(0)-\frac{4}{3}\lambda_{n,k}^{3}(0)
\end{equation*}
for $\left|c_{n,k}(M)\right|\geq 1/3$.
\end{corollary}
\begin{proof}
For simplicity in notation we omit in the following the indices $n$
and $k$ of $\lambda$. When $\omega=0$ (\ref{ODE}) reduces to
\[
\fl \frac{d\lambda_{n,k}}{dM}=-\frac{2m_e k}{4\lambda^2-1}.
\]
Taking into account that the above ODE is separable, the computation
of
\[
\fl
\int_{\lambda(0)}^{\lambda(M)}\,\rmd\lambda~\left(4\lambda^2-1\right)=-2m_e
k\int_{0}^{M}\,\rmd M
\]
gives rise to the following cubic equation for $\lambda$
\[
\fl \frac{4}{3}\lambda^3(M)-\lambda(M)+c=0,\qquad
c:=2kMm_e+\lambda(0)-\frac{4}{3}\lambda^3(0).
\]
Since $c\in\mathbb{R}$ the only real root of the above equation is
\[
\fl
\lambda=\frac{1+\left(-3c+\sqrt{9c^2-1}\right)^{2/3}}{2\sqrt[3]{-3c+\sqrt{9c^2-1}}}
\]
for $9c^2\geq 1$.\hspace{0.5cm}$\square$
\end{proof}

\section{A quasi-linear PDE for the eigenvalues $\lambda$}\label{sec:4}
Analogously to \cite{BA} we can study the eigenvalues of the angular
problem as a function of the parameters $\mu:=Mm_e$ and
$\nu:=M\omega$. Since the procedure is the same as in \cite{BA} with
the only exception that now $a=M$ where $a$ is the angular momentum
per unit mass of the black hole we limit us to present the main
result.
\begin{theorem}\label{PDE}
For fixed $k$ and $n$ the n-th eigenvalue
$\lambda=\lambda_n(k;\mu,\nu)$ is an analytical function in
$(\mu,\nu)\in\mathbb{R}^2$ satisfying the first order quasi-linear
partial differential equation
\begin{equation}\label{PDE1}
\fl
(\mu+2\nu\lambda)\frac{\partial\lambda}{\partial\mu}+(\nu+2\mu\lambda)\frac{\partial\lambda}{\partial\nu}+2k\mu-2\mu\nu=0
\end{equation}
with $\lambda_n(k,0,0)$ given by (\ref{initial_condition}).
\end{theorem}
\begin{proof}
Same proof as in Thm.1, Sec.III in \cite{BA} with $\nu$ replaced now
by $-\nu$.\hspace{0.5cm}$\square$
\end{proof}
In order to derive formal power series solutions of (\ref{PDE1}) it
is convenient to introduce a new function $\Lambda(\mu,\nu)$ defined
by the relation $\lambda(\mu,\nu)=\Lambda(\mu,\nu)+\lambda_n(k,0,0)$
and new independent variables $\widetilde{\mu}:=\mu-\nu$ and
$\widetilde{\nu}:=\mu+\nu$. Hence, (\ref{PDE1}) becomes
\begin{equation}\label{PDE2}
\fl
a_{1}(\widetilde{\mu},\widetilde{\nu},\Lambda)\frac{\partial\Lambda}{\partial\widetilde{\mu}}+
a_{2}(\widetilde{\mu},\widetilde{\nu},\Lambda)\frac{\partial\Lambda}{\partial\widetilde{\nu}}=f(\widetilde{\mu},\widetilde{\nu})
\end{equation}
with
\begin{equation}\label{coefficientfunctions}
\fl
a_{1}(\widetilde{\mu},\widetilde{\nu},\Lambda):=\widetilde{\mu}\left[1-2(\Lambda+\lambda_n(k,0,0))\right],\quad
a_{2}(\widetilde{\mu},\widetilde{\nu},\Lambda):=\widetilde{\nu}\left[1+2(\Lambda+\lambda_n(k,0,0))\right]
\end{equation}
and
\begin{equation}\label{source}
\fl
f(\widetilde{\mu},\widetilde{\nu}):=\frac{1}{2}\left(\widetilde{\nu}^{2}-\widetilde{\mu}^{2}\right)-
k\left(\widetilde{\nu}+\widetilde{\mu}\right).
\end{equation}
Since $\Lambda(0,0)=0$, $a_{i}(0,0,0)=0$ for all $i=1,2$ and
$f(0,0)=0$ we can apply a method similar to that developed in
\cite{MASA} to study formal power series solutions of (\ref{PDE2}).
In what follows we are interested in the existence and uniqueness of
the formal solution
\begin{equation}\label{FPS}
\fl \Lambda(x)=\sum_{|\alpha|\geq 1}\Lambda_{\alpha}x^{\alpha},\quad
\alpha=(\mathfrak{m},\mathfrak{n})\in\mathbb{N}^{2},\quad
|\alpha|=\mathfrak{m}+\mathfrak{n}, \quad
x=(\widetilde{\mu},\widetilde{\nu})\in\mathbb{R}^{2}
\end{equation}
centered at the origin for the equation (\ref{PDE2}). Moreover, we
will investigate the convergence of the formal power series
solutions (\ref{FPS}) by computing its Gevrey order. We recall that
a function $F(x)$ with $x=(x_1,x_2)\in\mathbb{R}^2$ is said to be of
Gevrey-$\{s\}$ class with $(s_1,s_2)\in\mathbb{R}^2$ if the power
series
\[
\fl
B_{s}\left[F\right](x)=\sum_{|\alpha|\in\mathbb{N}^{2}}F_{\alpha}\frac{x^{\alpha}}{\left(\alpha!\right)^{s-1^{(2)}}},
\quad 1^{(2)}:=(1,1),\quad
\left(\alpha!\right)^{s-1^{(2)}}:=(\mathfrak{m}!)^{s_1-1}(\mathfrak{n}!)^{s_2-1}
\]
converges in a neighborhood of $x=0$. By $G^{\left\{s\right\}}$ we denote the set of all formal power series of Gevrey-$\{s\}$ class. Furthermore, $F(x)\in G^{\left\{(1,1)\right\}}$ if and only if $F(x)$ is a convergent power series near $x=0$. For further details we refer to \cite{BAL}.\\
Let $\mathfrak{J}$ be the Jacobi matrix of the vector field
\[
\fl
x\longmapsto\left(a_1(x,0),a_2(x,0)\right)=\left(\widetilde{\mu}(1-2\lambda_n(k,0,0)),\widetilde{\nu}(1+2\lambda_n(k,0,0))\right)
\]
with $x$ defined by (\ref{FPS}). Then, we have
\begin{equation}\label{jacobi}
\fl \mathfrak{J}=\left(\left.\frac{\partial a_i(x,0)}{\partial
x_j}\right|_{x=0}\right)_{i,j=1,2}= \left( \begin{array}{cc}
                            1-2\lambda_n(k,0,0)&0\\
                            0&1+2\lambda_n(k,0,0)
                            \end{array} \right).
\end{equation}
\begin{lemma}\label{determinant}
For all $k=\pm\frac{1}{2},\pm\frac{3}{2},\cdots$ and
$n\in\mathbb{Z}\backslash\{0\}$ it results $\rm{det}\mathfrak{J}\neq
0$.
\end{lemma}
\begin{proof}
An elementary computation involving (\ref{initial_condition}) gives
\[
\fl
\mbox{det}\mathfrak{J}=1-4\lambda_n^2(k,0,0)=1-4\left(|k|-\frac{1}{2}+|n|\right)^2.
\]
Since $|k|\geq 1/2$ and $|n|\geq 1$ it follows that
$|k|-\frac{1}{2}+|n|\geq 1$. Hence, $\mbox{det}\mathfrak{J}\leq
-3$.\hspace{0.5cm}$\square$
\end{proof}
In the next lemma we show that the so-called
Poincar$\acute{\mbox{e}}$ condition is satisfied by (\ref{PDE2}).
\begin{lemma}\label{Poincare}
For all $\alpha\in\mathbb{N}^{2}$ it results
\begin{equation}\label{Poin_condition}
\fl
\left|\lambda_1\mathfrak{m}+\lambda_2\mathfrak{n}-f_\Lambda(0)\right|>c|\alpha|,\quad
f_\Lambda(0):=\left.\frac{\partial f}{\partial\Lambda}\right|_{x=0}
\end{equation}
where $\lambda_1$ and $\lambda_2$ denote the eigenvalues of the
Jacobi matrix $\mathfrak{J}$, $f$ is given by (\ref{source}) and $c$
is a positive constant independent of $\alpha\in\mathbb{N}^{2}$.
\end{lemma}
\begin{proof}
Taking into account that $\partial f/\partial\Lambda=0$ and
employing (\ref{jacobi}) we obtain
\begin{equation}\label{prima}
\fl
\left|\lambda_1\mathfrak{m}+\lambda_2\mathfrak{n}-f_\Lambda(0)\right|^2=(1-2\lambda_{n,k}(0))^2\mathfrak{m}^2+
(1+2\lambda_{n,k}(0))^2\mathfrak{n}^2+2(1-4\lambda_{n,k}^2(0))\mathfrak{m}\mathfrak{n}
\end{equation}
where $\lambda_{n,k}(0)=\lambda_n(k,0,0)$. To prove
(\ref{Poin_condition}) we have to distinguish between the cases
$n>0$ and $n<0$. In what follows we give the proof for $n>0$ since
the case $n<0$ can be treated analogously. Let us rewrite
(\ref{initial_condition}) as
$\lambda_{n,k}(0)=\rm{sign}(n)|\lambda_{n,k}(0)|$ with
$\lambda_{n,k}(0)=|k|-\frac{1}{2}+|n|$. For $n>0$ (\ref{prima})
becomes
\begin{equation*}
\fl
\left|\lambda_1\mathfrak{m}+\lambda_2\mathfrak{n}-f_\Lambda(0)\right|^2=(1-2|\lambda_{n,k}(0)|)^2\mathfrak{m}^2+
(1+2|\lambda_{n,k}(0)|)^2\mathfrak{n}^2+2(1-4\lambda_{n,k}^2(0))\mathfrak{m}\mathfrak{n}.
\end{equation*}
Since $1+2|\lambda_{n,k}(0)>1-2|\lambda_{n,k}(0)$ for all
$n\in\mathbb{N}\backslash\{0\}$ and $k=\pm 1/2,\pm 3/2,\cdots$ the
above expression can be majorized as
\[
\fl
\left|\lambda_1\mathfrak{m}+\lambda_2\mathfrak{n}-f_\Lambda(0)\right|^2>|1-2|\lambda_{n,k}(0)||^2(\mathfrak{m}+
\mathfrak{n})^2.
\]
Hence, (\ref{Poin_condition}) is satisfied for some positive
constant $c:=|1-2|\lambda_{n,k}(0)||$. \hspace{0.5cm}$\square$
\end{proof}
Finally, notice that
\begin{equation}\label{last_condition}
\fl \left.\frac{\partial
a_{i}}{\partial\Lambda}(\widetilde{\mu},\widetilde{\nu},\Lambda)\right|_{(0,0,0)}=0\quad\mbox{for
all $i=1,2$}.
\end{equation}
Since the Poincar$\acute{\mbox{e}}$ condition (\ref{Poin_condition})
and (\ref{last_condition}) are satisfied Thms.$1.1-2$, $\S$$1.2$ in
\cite{MASA} imply that the equation (\ref{PDE2}) has a unique formal
power solution
\begin{equation}\label{serie}
\fl \Lambda(\widetilde{\mu},\widetilde{\nu})=\sum_{|\alpha|\geq
1}\Lambda_{\alpha}~\widetilde{\mu}~^{\mathfrak{m}}\widetilde{\nu}~^{\mathfrak{n}},\quad
\alpha=(\mathfrak{m},\mathfrak{n})\in\mathbb{N}^{2},\quad
|\alpha|=\mathfrak{m}+\mathfrak{n}.
\end{equation}
Furthermore, since $\Lambda\in G^{\{(1,1)\}}$ it results that
(\ref{serie}) converges. The last task is to compute a recurrence
relation for the coefficients $\Lambda_\alpha$. To this purpose we
rewrite (\ref{serie}) as follows
\begin{equation}\label{serie1}
\fl
\Lambda(\widetilde{\mu},\widetilde{\nu})=\sum_{\mathfrak{m},\mathfrak{n}=0}^{\infty}
\Lambda_{\mathfrak{m},\mathfrak{n}}~\widetilde{\mu}~^{\mathfrak{m}}\widetilde{\nu}~^{\mathfrak{n}},\quad
\mathfrak{m}+\mathfrak{n}\geq 1.
\end{equation}
Since
\[
\fl
\Lambda(\widetilde{\mu},\widetilde{\nu})^2=\sum_{\mathfrak{m},\mathfrak{n}=0}^{\infty}
\left(\sum_{\mathfrak{r}=0}^{\mathfrak{m}}\sum_{\mathfrak{s}=0}^{\mathfrak{n}}
\Lambda_{\mathfrak{r},\mathfrak{s}}\Lambda_{\mathfrak{m}-\mathfrak{r},\mathfrak{n}-\mathfrak{s}}
\right)\widetilde{\mu}~^{\mathfrak{m}}\widetilde{\nu}~^{\mathfrak{n}},\quad
\mathfrak{m}+\mathfrak{n}\geq 1,\quad \mathfrak{r}+\mathfrak{s}\geq
1
\]
from (\ref{PDE2}) we obtain the identity
\[
\fl
\sum_{\mathfrak{m},\mathfrak{n}=0}^{\infty}\left((\mathfrak{m}c_{-}+\mathfrak{n}c_{+})\Lambda_{\mathfrak{m},\mathfrak{n}}
+(\mathfrak{n}-\mathfrak{m})\sum_{\mathfrak{r}=0}^{\mathfrak{m}}\sum_{\mathfrak{s}=0}^{\mathfrak{n}}
\Lambda_{\mathfrak{r},\mathfrak{s}}\Lambda_{\mathfrak{m}-\mathfrak{r},\mathfrak{n}-\mathfrak{s}}\right)
\widetilde{\mu}~^{\mathfrak{m}}\widetilde{\nu}~^{\mathfrak{n}}=
\frac{1}{2}\left(\widetilde{\nu}^{2}-\widetilde{\mu}^{2}\right)-
k\left(\widetilde{\nu}+\widetilde{\mu}\right)
\]
with $\mathfrak{m}+\mathfrak{n}\geq 1$ and $c_{\pm}:=1\pm
2\lambda_{n}(k,0,0)$. If we compare the terms of equal order in
$\widetilde{\mu}$ and $\widetilde{\nu}$ it follows that
\[
\fl \Lambda_{1,0}=\frac{k}{2\lambda_{n}(k,0,0)-1},\quad
\Lambda_{0,1}=-\frac{k}{2\lambda_{n}(k,0,0)+1},\quad\Lambda_{11}=0
\]
\[
\fl
\Lambda_{2,0}=\frac{(2\lambda_{n}(k,0,0)-1)^2-4k^2}{4(2\lambda_{n}(k,0,0)-1)^3},\quad
\Lambda_{0,2}=\frac{(2\lambda_{n}(k,0,0)+1)^2-4k^2}{4(2\lambda_{n}(k,0,0)+1)^3},
\]
and for $\mathfrak{m}+\mathfrak{n}\geq 3$ the coefficients
$\Lambda_{\mathfrak{m},\mathfrak{n}}$ are given by the relation
\begin{equation}\label{recurrence}
\fl
\Lambda_{\mathfrak{m},\mathfrak{n}}=\frac{\mathfrak{m}-\mathfrak{n}}{\mathfrak{m}c_{-}+\mathfrak{n}c_{+}}
\sum_{\mathfrak{r}=0}^{\mathfrak{m}}\sum_{\mathfrak{s}=0}^{\mathfrak{n}}
\Lambda_{\mathfrak{r},\mathfrak{s}}\Lambda_{\mathfrak{m}-\mathfrak{r},\mathfrak{n}-\mathfrak{s}},\quad
\mathfrak{r}+\mathfrak{s}\geq 1.
\end{equation}
Finally, notice that for $\mathfrak{m}=\mathfrak{n}$ we have
$\Lambda_{\mathfrak{n},\mathfrak{n}}=0$ for all
$\mathfrak{n}\in\mathbb{N}$.

\section{Analysis of the radial system (\ref{radial})}\label{sec:5}
In this section we show that there exists no bound state for the
Dirac equation in the extreme Kerr metric. The main idea behind the
proof is that after a suitable transformation the deficiency indices
of the transformed radial operator are zero. In fact, the deficiency
index of a differential operator simply counts the number of square
integrable solutions.
\begin{theorem}
The solution set of the system (\ref{cond1}) with (\ref{cond2}) or
(\ref{cond3}) is empty.
\end{theorem}
\begin{proof}
In order to apply some results of \cite{las} we bring the radial
system (\ref{radial}) in a more amenable form by transforming the
dependent variable as $f(r)=(f_1(r),f_2(r))^t:=(F(r)-\rmi
G(r),F(r)+\rmi G(r))^t$. Moreover, we introduce a new independent
variable $x$ defined by the relation
\[
\fl \frac{dx}{dr}=\frac{r^2+M^2}{(r-M)^2}.
\]
The solution of the above equation is
\[
\fl x(r)=r-\frac{2M^2}{r-M}+2M\ln{(r-M)}
\]
and it can be easily seen that $x\in\mathbb{R}$ since $x\to+\infty$
for $r\to+\infty$ and $x\to-\infty$ for $r\to M^+$. Let
$\Xi=(F,G)^t$. Hence, (\ref{radial}) becomes
\begin{equation}\label{system1}
\fl (\mathfrak{U}~\Xi)(x)=J~\frac{d\Xi}{dx}+B(x)~\Xi=\omega~\Xi
\end{equation}
with
\[
\fl J=\left( \begin{array}{cc}
                            0&+1\\
                            -1&0
                             \end{array} \right),\qquad
B(x)=\left( \begin{array}{cc}
                         -\frac{m_{e}r(x)(r(x)-M)+kM}{r^2(x)+M^2}&\lambda\frac{r(x)-M}{r^2(x)+M^2}\\
                          \lambda\frac{r(x)-M}{r^2(x)+M^2}&\frac{m_{e}r(x)(r(x)-M)-kM}{r^2(x)+M^2}
                             \end{array} \right).
\]
According to the above transformations the integrability condition
(\ref{integrability_cond}) for the radial spinors  simplifies to
\[
\fl (\Xi,\Xi)=\int_{-\infty}^{+\infty}\,\rmd x~
(F^2(x)+G^2(x))<\infty.
\]
Notice that the formal differential expression $\mathfrak{U}$ is
formally symmetric since $J=-J^{*}$ and $B=B^{*}$. Let
$\mathcal{S}_{min}$ be the minimal operator associated to
$\mathfrak{U}$ such that $\mathcal{S}_{min}$ acts in the Hilbert
space $L^{2}(\mathbb{R},dx)^2$ with respect to the scalar product
$(\cdot,\cdot)$. The operator $\mathcal{S}_{min}$ with domain of
definition $D(\mathcal{S}_{min})=C_{0}^{\infty}(\mathbb{R})^2$ such
that $\mathcal{S}_{min}~\Xi:=\mathfrak{U}~\Xi$ for $\Xi\in
D(\mathcal{S}_{min})$ is densely defined and closable. Let $S$
denote the closure of $\mathcal{S}_{min}$. We apply the so-called
decomposition method due to Neumark \cite{NEUM}. To this purpose,
let $\mathcal{S}_{min,\pm}$ be the minimal operators associated to
$\mathfrak{U}$ when restricted on the half-lines $[0,\infty)$ and
$(-\infty,0]$, respectively. We consider $\mathcal{S}_{min,\pm}$
acting in the Hilbert spaces $L^{2}(\mathbb{R}_{\pm},dx)^2$ with
respect to the scalar product $(\cdot,\cdot)$. The operators
$\mathcal{S}_{min,\pm}$ given by
$D(\mathcal{S}_{min,\pm})=C_{0}^{\infty}(\mathbb{R}_{\pm})^2$ with
$\mathcal{S}_{min,\pm}~\Xi_{\pm}:=\mathfrak{U}~\Xi_{\pm}$ for
$\Xi_{\pm}\in D(\mathcal{S}_{min,\pm})$ are densely defined and
closable. Notice that since the formal differential operator
$\mathfrak{U}$ is in the limit point case at $\pm\infty$ the
operators $\mathcal{S}_{min,\pm}$ are even essentially self-adjoint.
In the following we denote the closure of $\mathcal{S}_{min,\pm}$ by
$S_{\pm}$. Let $N_{\pm}(S_{\pm})$ be the deficiency indices of the
system (\ref{system1}) and let us denote by $\kappa_{\pm}$ the
number of positive and negative eigenvalues of the matrix $\rmi J$.
Clearly, we have $\kappa_+=1=\kappa_{-}$. Thm~5.2 (see Sect.~5.1 in
\cite{las}) implies that $N_{\pm}(S_{+})=1=N_{\pm}(S_{-})$.
According to Def.~2.14 (Sect.~2.5 in \cite{las}) the system
(\ref{system1}) is definite on $\mathbb{R}_+$ and on $\mathbb{R}_-$
and Prop.~5.4 (Sect.~5.1 in \cite{las}) implies that the deficiency
indices for $S$ are
\[
\fl N_{\pm}(S)=N_{\pm}(S_{+})+N_{\pm}(S_{-})-2=0.
\]
Therefore, the system (\ref{system1}) does not admit any square
integrable solution on the whole real line and this completes the
proof. \hspace{0.2cm}$\square$
\end{proof}
To conclude, we mention that in \cite{MO} a set of necessary and
sufficient conditions for the existence of an energy eigenvalue for
the Dirac equation in the extreme Kerr-Newman metric have been
derived. However, also in that case the conditions are so
complicated that it is not clear at all if they admit at least one
non-trivial solution. We believe that the approach we used to show
the absence of bound state solutions for the Dirac equation in the
extreme Kerr metric should work as well in the case of the extreme
Kerr-Newman metric since in the latter case the radial system can be
written again a form similar to (\ref{system1}).

\end{document}